\documentclass[twocolumn,english,prl,showpacs]{revtex4-1}

\usepackage{float}
\usepackage[dvips,final]{graphicx}
\usepackage{color}
\usepackage{amsmath}
\usepackage[T1]{fontenc}
\usepackage{relsize}
\usepackage{bm}
\usepackage{verbatim}

\usepackage{appendix}
\usepackage{booktabs}
\usepackage{xcolor}
\usepackage{soul}
\usepackage{natbib}
\usepackage{ulem}

\newcommand{\prlsection}[1]{\textbf{ #1 }~\textemdash~}

\begin{document}
\title{Probing Iron in Earth's Core With Molecular-Spin Dynamics}

\author{S. Nikolov$^1$}
\email{Corresponding author: snikolo@sandia.gov}
\author{K. Ramakrishna$^{2,3}$}
\author{A. Rohskopf$^1$}
\author{M. Lokamani$^3$}
\author{J. Tranchida$^4$}
\author{J. Carpenter$^1$}
\author{A. Cangi$^{2,3}$}
\author{M.A. Wood$^1$}

\affiliation{$^1$ Computational Multiscale Department, Sandia National Laboratories, Albuquerque, NM, United States}
\affiliation{$^2$ Center for Advanced Systems Understanding (CASUS), D-02826 G\"orlitz, Germany}
\affiliation{$^3$ Helmholtz-Zentrum Dresden-Rossendorf (HZDR), D-01328 Dresden, Germany}
\affiliation{$^4$ CEA, DES/IRESNE/DEC, 13018 Saint Paul Lès Durance, France}

\date{\today}

\begin{abstract}

Dynamic compression of iron to Earth-core conditions is one of the few ways to gather important elastic and transport properties needed to uncover key mechanisms surrounding the geodynamo effect. 
Herein a new machine-learned ab-initio derived molecular-spin dynamics (MSD) methodology with explicit treatment for longitudinal spin-fluctuations is utilized to probe the dynamic phase-diagram of iron. 
This framework uniquely enables an accurate resolution of the phase-transition kinetics and Earth-core elastic properties, as highlighted by compressional wave velocity and adiabatic bulk moduli measurements.
In addition, a unique coupling of MSD with time-dependent density functional theory enables gauging electronic transport properties, critically important for resolving geodynamo dynamics.

\end{abstract}
\pagebreak
\maketitle

\prlsection{Introduction}
Iron, Earth's most abundant element by mass, unveils intricate and multifaceted behaviors under extreme temperatures and pressures. For millennia, iron has been an integral part of human civilization, with approximately 90\% of global metal refining dedicated to its myriad of applications, spanning from kitchen utensils and structural building components, to microscopic drug delivery systems~\cite{dagdelen2023redox,denmark2016remote}.
The significance of iron transcends its uses on and above the terrestrial surface. Earth's core, predominantly constituted of iron, assumes a central role in the planet's geophysical and geochemical processes~\cite{CW02}. At pressures approaching 350 GPa and temperatures surpassing 6000 K, the flow of iron in the planet's core acts as the source of Earth's magnetic field, shielding our planet from detrimental solar radiation while exerting influence over plate tectonics and mantle convection. Hence, it is not an overstatement that iron is critical for life on Earth. Yet still, the precise mechanisms responsible for upholding Earth's magnetic fields remain unclear~\cite{LFNC22}. The prevailing theory posits a dynamo mechanism within the outer core, where the convective motion of molten iron generates electrical currents, giving rise to the magnetic field we observe~\cite{F73, L03, SL07}.
Unraveling the geodynamo mechanism crucially depends on the underlying material properties. Understanding the phase diagram of iron, especially its melting line~\cite{KHAB22}, and transport properties, such as electrical and thermal conductivity, across a wide spectrum of temperatures and pressures, are pivotal in this pursuit~\cite{PDGA12}.

Numerous experimental endeavors have significantly enhanced our comprehension of the iron phase diagram. Early measurements employed both compressive~\cite{BPM56} and shock~\cite{A86} waves to investigate polymorphic phases at low pressure and the phase diagram up to  Earth-core conditions. 
Recent measurements~\cite{TB64,MBIP04,DLOM06,KMNH20, HGCE20, WKVL20, GASP21}, utilizing short pulse optical lasers, laser-driven shocks, and dynamic compression techniques, have further expanded our knowledge of iron's phase diagram. Although fewer in number, investigations of iron's transport properties have been conducted using DAC~\cite{GOHL13,OKHS16,KMGG16,OSKO23}, wire-heating~\cite{KMGG16,BPJ94}, and both static and dynamic shock-compression~\cite{KM69,G83,G86}. Remarkably, experiments involving laser-heated DAC~\cite{OKHS16,KMGG16} have ignited controversy in measuring electronic transport properties at Earth-core conditions~\cite{D16}. These indispensable experimental efforts demand significant effort to achieve the necessary accuracy, and therefore motivate computational efforts to fill in gaps in transport properties throughout the phase diagram.

\begin{figure*}[htp]
\centering
\includegraphics[width=\textwidth]{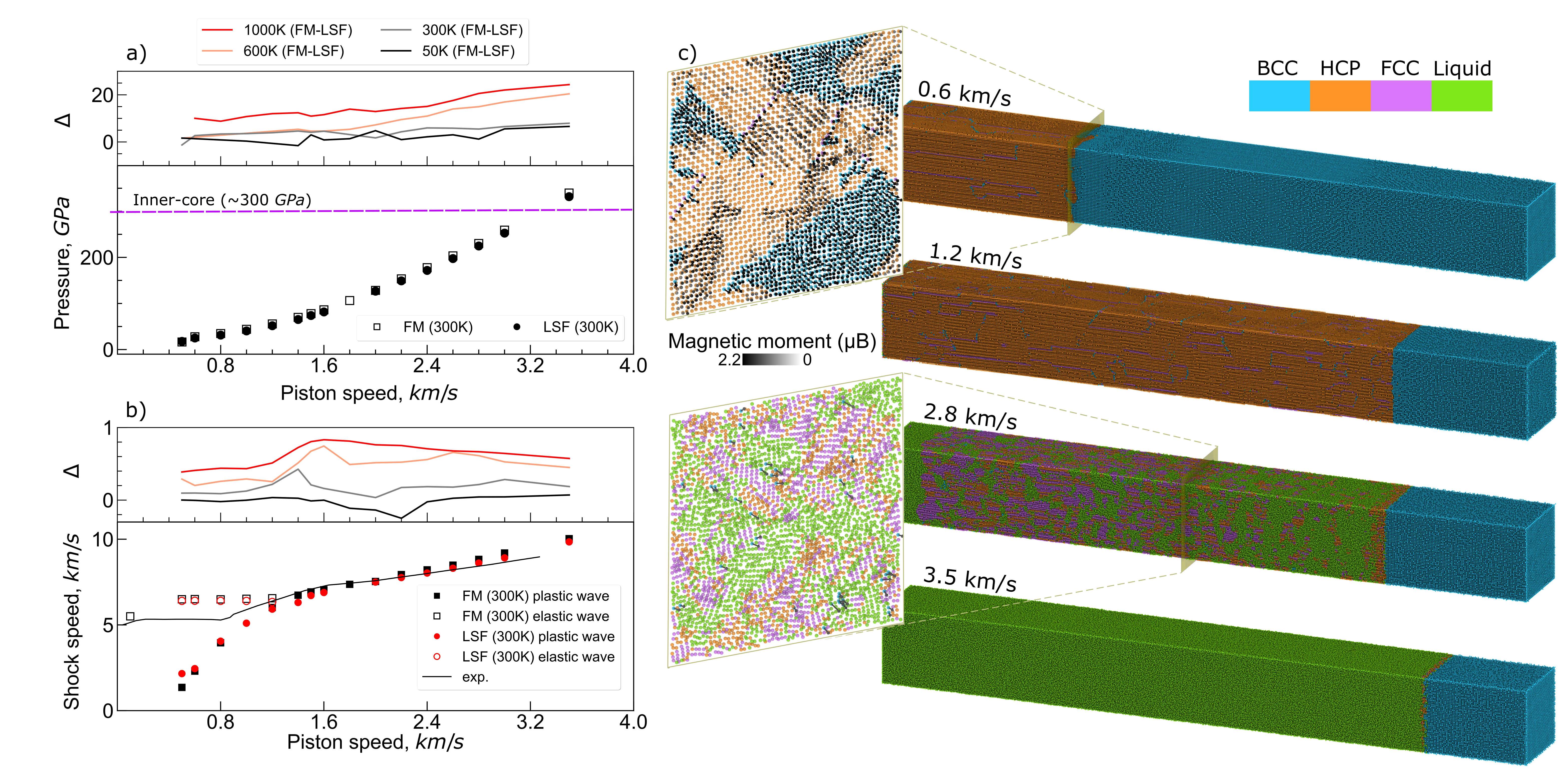}
\caption{a) Hugoniot states ($u_{p}$, P) in shocked iron for both FM and LSF cases. The top inset is the pressure difference ($\Delta$) between FM-LSF calculations, with a common axis along $u_{p}$. b) Hugoniot states ($u_{p}$, $u_{s}$) for an initial temperature of 300$K$. Upper panel captures $\Delta=u_{s}^{FM}-u_{s}^{LSF}$. For $u_{p}\leq$ 1.2 km/s both elastic and plastic wavefronts are separated spatially, above which a single wave is observed. c) MSD snapshots from LSF simulations at different piston speeds, color indicating phase, and grayscale arrows for spin vectors where present (see magnetic moment legend). At $u_{p}\simeq0.6$ km/s a clear BCC-HCP transition is detected where magnetic moments for HCP atoms decrease rapidly with pressure, see inset for clarity. The onset of melting occurs $u_{p}\geq2.8$ km/s, lower close-up image shows few atoms retain their magnetic moments in this molten state.} 
\label{fig:landscapes}
\end{figure*}

Previous simulation efforts, employing various levels of theory, have extensively explored the iron phase diagram, encompassing its polymorphs and phases at high pressure and temperature~\cite{HP83, P19, KYPO21}. Furthermore, there has been a growing emphasis on the development of increasingly accurate equation of state models~\cite{ZLPH10, SC10, DDLS17}. More recently, classical molecular dynamics (MD) simulations~\cite{AW59}, based on the parameterization of a Born-Oppenheimer potential energy surface (BO-PES) \cite{R04} using machine-learning techniques, have emerged as the state of the art in atomistic modeling. A substantial number of highly precise machine-learned interatomic potentials have been built from ab-initio calculations~\cite{KS65,BP07, HBCK17, SIR17, ZHWC18, BPKC10, JNG14, LSB18, SSKT18, TSTF15} with various model forms and feature sets. While invaluable for large-scale investigations of material properties~\cite{LCZZ20, CWT20}, these new potentials have primarily been confined to non-magnetic phases, consequently making inaccurate predictions of simple properties like heat capacity~\cite{DDCM18}. To resolve this, recent efforts have incorporated atomic spin dynamics into MD simulations~\cite{MWD08,TPTT18}. Although early models provided qualitative agreement with experimental results~\cite{MDW16, DALM20, ZTGM20} via embedded-atom-method potentials~\cite{MD20}, recent advancements have successfully constructed magneto-elastic potentials for coupled molecular-spin dynamics (MSD) simulations, achieving accuracies similar to the underlying first-principles methods~\cite{NWCM21, NTRL22, nieves2022atomistic, nikolov2023temperature,rohskopf2023fitsnap}.

In this letter, building on prior spin-dynamical efforts~\cite{ruban2007temperature,ma2012longitudinal,gambino2020longitudinal}, we incorporate longitudinal spin fluctuations (LSFs) into our MSD framework and demonstrate this necessary adaptation for the behavior of iron up to Earth-core conditions. To resolve magnon-phonon interactions within our computationally efficient MSD framework, a quantum-accurate ML potential is constructed that properly partitions magnetic/non-magnetic contributions to the BO-PES~\cite{NWCM21}. Due to experimental parallels, to study Earth-core states of matter, we simulate the iron single-crystalline Hugoniot curves up to Earth-core conditions, where changes in initial preheat temperature permit mapping out a large part of the structural polymorphs and liquidus. One of the unique benefits of the current MSD approach is the access to orders of magnitude larger, compared to DFT methods, temporal/spatial domains which minimizes the barrier to providing fundamental transport properties for geophysical models. To illustrate this capability, elastic properties of iron for a 5882$K$ isotherm are measured up to roughly 300 GPa, enabling direct comparison to preliminary reference Earth model (PREM) measurements for compressional "P-wave" velocities and adiabatic bulk modulus ($B$). Additional transport property measurements are enabled by a top-down multiscale modeling approach wherein from large MSD simulations representative structures can be isolated and used for detailed time-dependent density functional theory (TDDFT) simulations~\cite{RLBV23,Ramakrishna2023iop}. Electrical-resistivity measurements are carried out in the range of 2000-4000$K$ at both 140 and 212 $GPa$, highlighting the impact of LSFs and showing good agreement with experimental measurements by Ohta and Zhang \textit{et al.}~\cite{OSKO23,OKHS16,ZHLZ20}. These demonstrations usher in new computational approaches for magnetic materials that have been absent, or under-resolved, for high-energy density states of matter. 

\prlsection{Results and Discussion}
The iron phase diagram~\cite{HP83, A86, P19, HGCE20, WKVL20, KYPO21} and its equation of state \cite{GCT11, A73, DDLS17, A86, P19, ZLPH10, DLOM06, SC10, GASP21, KMNH20, GM23} have been thoroughly investigated in recent decades. 
Within these efforts the behavior of iron, particularly across large regions of temperature-pressure space has been difficult to pin down, partly due to the variety of phases iron can take on ($\alpha$, $\gamma$, $\epsilon$, $\delta$, liquid) and the underlying magnetic character for some of these phases. The ability to accurately resolve large portions of the iron phase diagram, in an atomistic setting, where one can predict, free from finite size/time restrictions the underlying grain structure, phase stability, transient (shock) kinetics, and transport properties (viscosity, electrical/thermal conductivity, self-diffusion, etc.) is of particular importance due to the underlying implications that these measurements have in the calculation of Earth-core~\cite{A86, MBT67, ADML13, KHAB22} properties. 

To this end, our simulation efforts begin by focusing on the construction of the iron phase diagram that includes a prediction of high-pressure/temperature metastable phases. 
An efficient means to sample large areas in $P$/$T$ phase space is to shock compress (Fig.~\ref{fig:landscapes}) a sample of varying initial (preheat) temperatures, where each piston impact velocity ($u_{p}$) generates a locus of Hugoniot points. Herein a key simulation capability advancement~\cite{luu2020shock,kadau2007shock} is the proper partitioning of magnetic and non-magnetic contributions to the potential energy surface, enabling a higher fidelity resolution of the transient shock dynamics. Longitudinal spin fluctuations may have previously been regarded as having a subtle effect on the ion dynamics, but as noted in Fig.~\ref{fig:landscapes} appreciable differences between the fixed-moment (FM - constant magnetic moment) and LSF cases can be seen. Namely, the top panels of Fig.~\ref{fig:landscapes}.a-b) compute the difference in pressure and shock velocity, respectively, where up to 25 $GPa$ and 1 $km/s$ differences can be seen for the shock simulations with the highest preheat.

\begin{figure*}[tp]
\centering
\includegraphics[width=\textwidth]{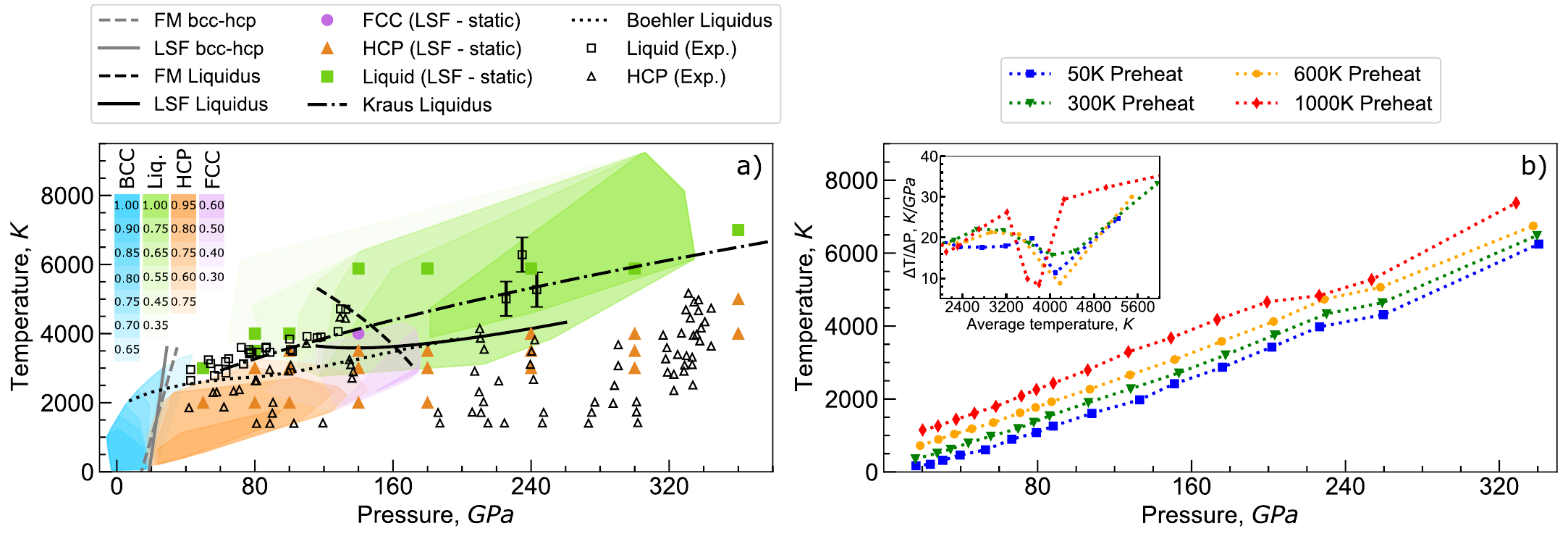}
\caption{a) Dynamic phase diagram for iron with MSD-LSF. Observed phases (colored areas) consist of four (50$K$, 300$K$, 600$K$, and 1000$K$) preheat temperatures~\cite{NH04}, which have been categorized according as light-blue - BCC, purple - FCC, orange - HCP, green - liquid. The liquid and HCP experimental data points are compiled from the works of Morard, Kuwayama, Anzellini, Jeffrey, Brown, Yoo, and Tateno \textit{et al.}~\cite{morard2014properties,kuwayama2020equation,tateno2010structure,NH04,yoo1995phase,brown1986phase,ADML13}. The predicted melt line from LSF simulations is in far better agreement with Kraus/Boehler~\cite{KHAB22,B93} lines than MSD-FM. b) Temperature vs. Pressure Hugoniot curves for the four different preheat temperatures. The inset shows how the numerical derivative of each curve varies with temperature, highlighting the predicted melting transition.}
\label{fig:phasediag}
\end{figure*}

The well-characterized \cite{kadau2007shock,jensen2009direct,hwang2020subnanosecond,luu2020shock} split elastic/plastic wave structure is recovered at $u_{p}<1.2$ $km/s$, with the $\alpha-\epsilon$ transition occurring at approx. $0.6$ $km/s$ ($\approx 16$ $GPa$), see top image of Fig.~\ref{fig:landscapes}.c.
In the MSD-LSF simulations (Fig.~\ref{fig:landscapes}.a,c), transitioning into the HCP phase brings about a strong decrease in the atomic spin magnitude which lowers the exchange/Landau contributions, resulting in lower pressure Hugoniot states. An in-depth overview of the LSF scheme and the corresponding strain-phase-magnetic-moment relationships can be found in the SI. Interestingly, both FM and LSF shock data predict a meta-stable FCC phase near the liquidus, $\sim160$ $GPa$ ($u_{p}\simeq2.2$ $km/s$).
The value in phase predictions from shock-compressed samples is this direct capture of meta-stability, which can help resolve differences in transport properties reported in the literature~\cite{OKHS16,KMGG16}.

Utilizing the MSD shock data here we construct a \textit{dynamic} phase diagram by sampling 10 $nm$ thick slices of material behind the shock front at 0.5 $ps$ intervals to identify $P$, $T$, and phases present. This data is collected in Fig.~\ref{fig:phasediag}.a. 
Color-shaded areas indicate MSD-LSF predicted phase fractions for BCC/HCP/FCC/Liquid (blue/orange/purple/green) phases, where overlapping regions indicate metastable mixed phases. Phase boundaries demarking $\alpha$-$\epsilon$ and liquidus for both LSF and FM are shown as lines and are constructed via support vector classification~\cite{DEFFRENNES2022110497,maulik2017remote}.
Additional liquidus lines from Boehler~\cite{B93} and Kraus~\cite{KHAB22} show much better agreement with LSF predictions than FM, highlighting the need for proper accounting of magnetic effects at high pressure and temperature conditions. Likewise for the BCC/HCP transition curves a difference between the FM/LSF case is observed, whereas for the LSF simulations, the BCC/HCP transition curve is shown to be less temperature sensitive.
 
\begin{figure}[th]
\centering
\includegraphics[width=1.0\columnwidth]{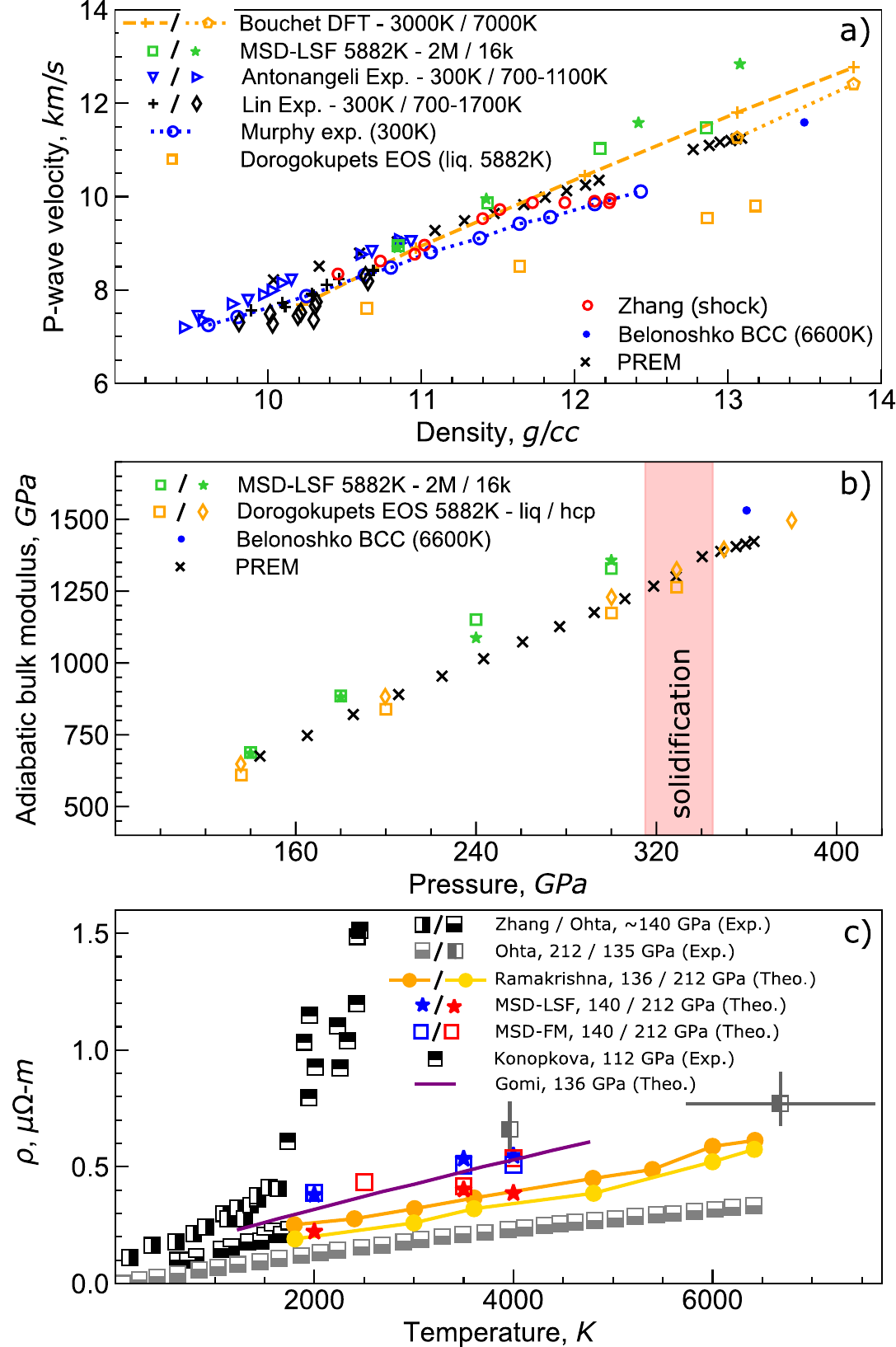}
\caption{a) Literature data~\cite{bouchet2022sound,antonangeli2012simultaneous,lin2005sound,murphy2013experimental} and predicted compressional wave MSD-LSF data (green markers) for liquid iron at $\sim$5882K. Finite-size effects are resolved with MSD, as exemplified by differences between 2M and 16k atom count simulations. b) Pressure-dependent adiabatic bulk modulus as compared to PREM, for 5882$K$ MSD-LSF isotherm, and 5882$K$ isotherms from Dorogokupets EOS model~\cite{DDLS17}. A lone data point for the BCC phase is included from Belonoshko \textit{et al.} to highlight the liquidus crossing near 320GPa~\cite{belonoshko2022elastic}. c) Temperature-dependent electrical resistivity for pressures close to 140 GPa and 212 GPa from experiment and theory. DAC measurements were reported by Ohta \textit{et al.}~\cite{OKHS16,OSKO23}, Zhang \textit{et al.}~\cite{ZHLZ20} and by Kon{\^o}pkov{\'a} \textit{et al.}~\cite{KMGG16}. Our calculations of the electrical conductivity using LSF/FM (red/blue stars and squares) are compared with previously reported calculations~\cite{GOHL13,RLBV23}.}
\label{fig:liquid_prop}
\end{figure}

Overlaying the dynamic phase diagram are static LSF-MSD calculations, filled symbols, which were performed using a large $2\cdot10^6$ atom simulation cell held at the designated $P/T$ for 200 $ps$. The static calculations here importantly fill in regions that are not well resolved within the dynamic phase diagram and agree well with experimental DAC results~\cite{kuwayama2020equation,morard2014properties,denoeud2016dynamic,yoo1995phase,morard2018solving,ADML13,tateno2010structure}. Notably within the static melt measurements, we do observe a metastable BCC transition (see Figure 5 in SI) which occurs right before the pair correlation function of the sample assumes the standard liquid profile, such metastability has been previously discussed by Bouchet and Belonoshko~\cite{belonoshko2021free,belonoshko2022elastic,bouchet2013ab}. Within the SI we provide a snapshot that shows how melting is initiated from the metastable BCC state.

An additional estimate of the liquidus, independent of the structure detection algorithm, is provided in Fig.~\ref{fig:phasediag}.b, which illustrates the LSF Hugoniot curves for each initial temperature used to construct our \textit{dynamic} phase diagram. The inflection in the P, T curve is indicative of a phase transition. The inset in Fig.~\ref{fig:phasediag}.b thus highlights the observed melt-transitions by showing how the derivatives of each curve vary with temperature. Based on this data the estimated melting point for a 300$K$ preheat is $[239 \mp 14GPa, \ 4182 \mp 123K]$. These results are in agreement with the previous laser multi-shock study by Ping \textit{et al.} which arrive at approx. $[250 \mp 30GPa, \ 4950 \mp 1050K]$~\cite{ping2013solid}. Meanwhile, sound velocity experiments by Nguyen \textit{et al.} find a melting point of approx $[226GPa, \ 5000 \mp 500K]$, which in terms of temperature, is a little bit higher than the current estimates. This agreement with the few available experimental studies gives support to the accuracy of the computational tools developed herein.

\textit{Elastic Properties} ---
Having constructed our \textit{dynamic} MSD-LSF phase diagram, we now turn to measurements of elastic and transport properties relevant to the proposed dynamo effect of Earth's core. Measurements for the compressional $P$-wave velocities and adiabatic bulk moduli for a 5882 $K$ isotherm within the pressure range of 140-300 $GPa$ were carried out for small ($1.6\cdot10^4$ atoms) and large ($2.0\cdot10^6$ atoms) liquid samples. This data is displayed in Fig.~\ref{fig:liquid_prop} along with available DFT, experimental, EOS, and PREM predictions. Note differences in temperature indicated in the legends as this will dictate the (meta-)stable phase present in the measurement. At the largest densities, we observe significant differences between the small and large MSD simulation cells which arise due to the fact that the large cells are all fully melted. In contrast, the smaller geometries retain a strong signal of BCC above 200 $GPa$ (additional details in SI).
The current findings thus support the existence of a high-pressure BCC phase, however, it is unclear whether this phase is a fully stable state or just a transient state that is encountered on the path to melting. 
While the current calculations point to the latter, a fully stable high-pressure BCC region may exist within an unresolved region of pressure/temperature space. 
Importantly, this has been hypothesized by Belonoshko  \textit{et al.} within their recent works~\cite{belonoshko2022elastic,belonoshko2021free,BLFZSS17}, as a potential way of explaining the anisotropy of Earth-core seismic wave measurements. 
Fig.~\ref{fig:liquid_prop}.b captures the current MSD-LSF results for the adiabatic bulk modulus and includes a corresponding comparison with PREM and EOS data for pure iron~\cite{DDLS17}. Again finite-size effects are observed between large and small simulation cells above 200 $GPa$. 

In general, we find that the largest deviations from the PREM are $\sim10\%$, showing the MSD-LSF model has great transferability over a large range of temperatures and pressures, something previously unachievable using standard MD approaches ~\cite{rosenbrock2021machine}. 
Note that experimental seismic measurements indicate the presence of lighter elements within the iron lattice~\cite{P94}. Recent studies have revealed that these light elements can impact elastic properties and melting temperatures at Earth-core pressures~\cite{he2022superionic,li2020shock,hirose2021light,UMEMOTO2020116009}. Thus an overly close agreement between PREM and pure iron results should not be anticipated. Solid BCC predictions from Belonoshko~\cite{belonoshko2022elastic} are also included in Fig.~\ref{fig:liquid_prop}.b, which albeit at a slightly higher temperature, agrees well with the trend in the MSD-LSF data. The red highlighted region here denotes where solidification is expected for the 5882 $K$ isotherm.

\textit{Transport Properties} ---
Lastly, we leverage the computational efficiency of MSD-LSF for ensemble sampling of structures which in turn are returned to \textit{ab-initio} codes for transport property measurements. The experiments with laser-heated DAC~\cite{OKHS16,KMGG16} have led to a notable controversy in the measurement of electronic transport properties in iron at the core-mantle boundary (CMB) and Earth-core conditions~\cite{D16,LG22}. 
Ohta \textit{et al.}~\cite{OKHS16} infer a thermal conductivity of 226~Wm$^{-1}$K$^{-1}$ by measuring the electrical resistance of iron wires and converting it into a thermal conductivity using the Wiedemann-Franz law~\cite{FW53}. 
Alternatively, Kon{\^o}pkov{\'a} \textit{et al.}~\cite{KMGG16} measured the thermal diffusion rate for heat transferred between the ends of solid iron samples, inferring a thermal conductivity of 33~Wm$^{-1}$K$^{-1}$ from the agreement with a finite-element model. 
The discrepancy in these measurements has deep implications for predicting the age of the Earth~\cite{D16}. 
Since the uncertainty in the electrical conductivity, both from experiment and theory is so high, reliable knowledge about the fundamental processes generating Earth's magnetic field is also lacking.
Due to the disagreement among existing experimental data, computational modeling is indispensable in supporting current and future efforts probing these properties~\cite{BS21}. 

Fig.~\ref{fig:liquid_prop}.c shows the electrical resistivity and its temperature dependence at a fixed pressure (P=140~GPa and P=212~GPa). Additional details regarding the simulation parameters can be found in the SI.
Structures sampled from LSF and FM MSD simulations are indicated by red/blue stars and squares respectively, highlighting the importance of proper magnetic treatment leading to these high-energy density states. The striking feature is that the predicted electrical resistivity agrees well with the recent measurements of Ohta \textit{et al.}~\cite{OSKO23}, particularly at 135~GPa and $\sim$4000~K as well as the earlier DAC measurements by Ohta and Zhang \textit{et al.} at 2000~K compared to other ab-initio models~\cite{GOHL13,RLBV23}. 
The influence of LSF on the ionic configurations results in a slightly higher resistivity at higher temperatures, especially between 3000~K to 4000~K in agreement with Ohta \textit{et al}. 

\prlsection{Conclusion}
The present work illustrates a promising simulation capability that incorporates an explicit treatment for magnetic exchange interactions and LSFs into efficient machine-learned interatomic potentials to resolve structural, mechanical, and transport properties of iron at high temperatures and pressures. 
Far-reaching effects of LSFs on key geophysical material properties were demonstrated, in particular phase stability in shock-compressed systems.
The prediction of a metastable FCC phase under shock near 160 GPa will test new experimental X-ray diagnostics at state-of-the-art light source facilities. 
Meanwhile, our static liquid calculations hint at a high-pressure BCC phase which could explain the anisotropic character of PREM seismic wave velocities. Recent studies have highlighted that a metastable inner-core BCC phase can be stabilized by the presence of light elements~\cite{WU2022,godwal2015stability}, and as such future work will leverage our unique MSD-TDDFT approach to determine the impact of this high-pressure BCC phase on electrical resistivities in Earth's inner core~\cite{WU2022,he2022superionic,godwal2015stability}. 

\prlsection{Acknowledgement}
This article has been authored by an employee of National Technology \& Engineering Solutions of Sandia, LLC under Contract No. DE-NA0003525 with the U.S. Department of Energy (DOE). The employee owns all right, title and interest in and to the article and is solely responsible for its contents. The United States Government retains and the publisher, by accepting the article for publication, acknowledges that the United States Government retains a non-exclusive, paid-up, irrevocable, world-wide license to publish or reproduce the published form of this article or allow others to do so, for United States Government purposes. The DOE will provide public access to these results of federally sponsored research in accordance with the DOE Public Access Plan \href{https://www.energy.gov/downloads/doe-public-access-plan}. KR and AC were supported by the Center for Advanced Systems Understanding (CASUS) which is financed by Germany’s Federal Ministry of Education and Research (BMBF) and by the Saxon state government out of the State budget approved by the Saxon State Parliament. Some computations were performed on a Bull Cluster at the Center for Information Services and High-Performance Computing (ZIH) at Technische Universit\"at Dresden, on the cluster Hemera at Helmholtz-Zentrum Dresden-Rossendorf (HZDR). We want to thank the ZIH for its support and generous allocations of computer time. 

\bibliography{main.bib}

\appendix
\end{document}


\title{Supplemental Material: Probing Iron in Earth's Core With Molecular-Spin Dynamics}

\author{S. Nikolov$^1$}
\author{K. Ramakrishna$^{2,3}$}
\author{A. Rohskopf$^1$}
\author{M. Lokamani$^3$}
\author{J. Tranchida$^4$}
\author{J. Carpenter$^1$}
\author{A. Cangi$^{2,3}$}
\author{M.A. Wood$^1$}

\affiliation{$^1$ Computational Multiscale Department, Sandia National Laboratories, Albuquerque, NM, United States}
\affiliation{$^2$ Center for Advanced Systems Understanding (CASUS), D-02826 G\"orlitz, Germany}
\affiliation{$^3$ Helmholtz-Zentrum Dresden-Rossendorf (HZDR), D-01328 Dresden, Germany}
\affiliation{$^4$ CEA, DES/IRESNE/DEC, 13018 Saint Paul Lès Durance, France}

\date{\today}
\pacs{31.15.A-,75.50.Ww, 75.30.Gw, 07.05.Tp }
\maketitle

\section{Molecular-spin dynamics}
 
All molecular-spin dynamics (MSD) calculations were carried out using the SPIN package~\cite{TPTT18} in LAMMPS~\cite{LAMMPS}, using a molecular-spin Hamiltonian that couples spin precession and atomic motion defined as:
\begin{equation}
\begin{aligned}
\mathcal{H}_{MSD}(\boldsymbol{r},\boldsymbol{p},\boldsymbol{s}) & =  \sum_{i=1}^N\frac{\boldsymbol{p}_i^2}{2m_i}+\sum_{i,j=1}^N\mathcal{V}^{SNAP}(r_{ij}) \\
& +\mathcal{H}_{ex}(\boldsymbol{r},\boldsymbol{s}) +\mathcal{H}_{Landau}(\boldsymbol{s})
,
\label{eq:Ham_tot}
\end{aligned}
\end{equation}
where $\boldsymbol{r}$, $\boldsymbol{p}$, and $\boldsymbol{s}$ are the per-atom position, momentum, and spin vectors, respectively. Meanwhile, $\mathcal{V}^{SNAP}$, $r_{ij}$, and $m_i$ are the SNAP interatomic potential, the separation distance between the $i-j$ particle pair, and the atomic mass, respectively. Spin exchange interactions $\mathcal{H}_{ex}$ use the biquadratic-exchange Hamiltonian, as shown in Eq.(~\ref{eq:exchange}). Functions ${J} \left(r_{ij} \right)$ and ${K} \left(r_{ij} \right)$ (in eV) are magnetic exchange functions, which are parameterized using the Bethe-Slater model form. For proper accounting of the magnetic pressure, the magnetic exchange function is offset by energy at the ferromagnetic ground state (0$K$) as detailed in Ma \textit{et al.}~\cite{MWD08}. The Landau expansion $\mathcal{H}_{Landau}$, describes how energy varies with changes in spin magnitude and is defined in Eq.~(\ref{HLandau}). Changes in the magnetic moment magnitude are applied using a strain-dependent first-principles parameterization for the $A$, $B$, and $C$ coefficients in Eq.~(\ref{HLandau})~\cite{gambino2020longitudinal}. 

\begin{eqnarray}
\mathcal{H}_{ex} &=& -\sum_{i\neq j}^{N} {J} \left(r_{ij} \right)\,
                      \left[\bm{s}_{i}\cdot \bm{s}_{j} -1 \right] \nonumber \\
                  &~& -\sum_{i\neq j}^{N} {K} \left(r_{ij} \right)\, 
                  \left[\left(\bm{s}_{i}\cdot \bm{s}_{j}\right)^2 -1 \right]\,
                  \label{eq:exchange}
\end{eqnarray}
\begin{equation}
\mathcal{H}_{Landau} = -\sum_{i\neq j}^{N} {A S_{i}^2 + B S_{i}^4 + C S_{i}^6} \
\label{HLandau}
\end{equation}

\noindent The equations of motion for an atomic particle within the classical MSD framework thus can be expressed as shown in Eqs.~\ref{SL_EOM_1}-\ref{SL_EOM_3}~\cite{P95,TPTT18,MD20}.
        
\begin{equation}
\begin{split}
\hspace{0.8mm} \frac{d\bm{r}_{i}}{dt} = \frac{\bm{p}_{i}}{m_i}
\label{SL_EOM_1}
\end{split}
\end{equation}

\begin{equation}
\begin{split}
\frac{d\bm{p}_{i}}{dt} &= \sum_{j,i \neq j}^{N} \bigg[ - \frac{dV_{SNAP}(r_{ij})}{dr_{ij}} + \frac{dJ(r_{ij})}{dr_{ij}}(\bm{s}_{i} \cdot \bm{s}_{j}) + ... \\ &\frac{dK(r_{ij})}{dr_{ij}}(\bm{s}_{i} \cdot \bm{s}_{j})^2\big] \bm{e}_{ij} 
\label{SL_EOM_2}
\end{split}
\end{equation}

\begin{equation}
\begin{split}
\hspace{0.4mm}  \frac{d\bm{s}_{i}}{dt} &= \frac{1}{(1+\lambda^2)}\bigg[(\bm{\omega}_{i}-\lambda\bm{\omega}_{i} \times \bm{s}_{i}\bigg] \times \bm{s}_{i}
\label{SL_EOM_3}
\end{split}
\end{equation}

 Here Eqs.~\ref{SL_EOM_1} and \ref{SL_EOM_2} describe the atomic equations of motion in the microcanonical ensemble, i.e. in the absence of a thermal or pressure bath. The Bethe-Slater model form~\cite{K92,YMY96} used to describe variables $J(r_{ij})$ and $K(r_{ij})$ is shown in Eq.~\ref{J_dep}. Equation~\ref{SL_EOM_3} represents the Landau-Lifshtiz-Gilbert equation, which defines the spin vector precessional motion, associated with each atom \cite{TPTT18}. The variable $\lambda$ is the transverse damping constant and $\bm{\omega}_i$ is the spin precessional vector~\cite{TPTT18}. 

\begin{equation}
  f\left(r \right) = 4\alpha
  \left(\frac{r}{\delta}\right)^2
  \left(1-\gamma\left(\frac{r}{\delta}\right)^2 \right)
  e^{-\left(\frac{r}{\delta}\right)^2}
  \Theta\left(R_c - r\right)\,
  \label{J_dep}
\end{equation}

Eqs.~\ref{SL_EOM_2}-\ref{SL_EOM_3} describe the shock-driven simulations where no thermo/barostats are applied, but these equations need to be extended to constrain the thermodynamic state to be studied. For preheating the samples, static phase-diagram calculations, and Born-matrix elastic property calculations these properties were sampled from the canonical or isobaric ensemble. Eqs.~\ref{SL_EOM_2}-\ref{SL_EOM_3} were modified to include the fluctuation-dissipation terms as shown in Eqs.~\ref{SL_EOM_4}-\ref{SL_EOM_5}.

\begin{equation}
\begin{split}
  \frac{d\bm{p}_{i}}{dt} &= \sum_{j,i \neq j}^{N} \bigg[ - \frac{dV_{SNAP}(r_{ij})}{dr_{ij}} + \frac{dJ(r_{ij})}{dr_{ij}}(\bm{s}_{i} \cdot \bm{s}_{j}) + ... \\ &\frac{dK(r_{ij}}{dr_{ij}}(\bm{s}_{i} \cdot \bm{s}_{j})^2\bigg] \bm{e}_{ij} - \frac{\gamma_{L}}{m_i}\bm{p}_{i} + \bm{f}(t)
\label{SL_EOM_4}
\end{split}
\end{equation}

\begin{equation}
\begin{split}
\hspace{0.4mm}  \frac{d\bm{s}_{i}}{dt} &= \frac{1}{(1+\lambda^2)}\bigg[(\bm{\omega}_{i}+\bm{\eta}(t)) \times \bm{s}_{i} + ...\\ &\lambda\bm{s}_{i} \times (\bm{\omega}_{i} \times \bm{s}_{i})\bigg]
\label{SL_EOM_5}
\end{split}
\end{equation}

Here, $\gamma_{L}$ and $\bm{f}$ are the Langevin dampening lattice parameter and Gaussian fluctuating force term, respectively, as defined in Eqs.~\ref{F-D_1}-\ref{F-D_2}. Within Eq.~\ref{F-D_2} the variable $T$ is the temperature, $k_B$ is the Boltzmann constant, and $\alpha$ and $\beta$ are coordinates. 

\begin{eqnarray}
  \langle \bm{f}(t) \rangle &=& 0 
\label{F-D_1}\\
  \langle f_{\alpha}(t) f_{\beta} (t') \rangle &=& 2 k_B T \gamma_L \delta_{\alpha \beta} \delta (t - t') 
\label{F-D_2}
\end{eqnarray}

Similarly to $\bm{f}$ the variable $\bm{\eta}$ is a Gaussian fluctuating force term with magnitude $D_{S} = 2\pi \lambda k_B T/\hbar$ ~\cite{MD20}.

\begin{eqnarray}
  \langle \bm{\eta}(t) \rangle &=& 0
\label{fd_spin_1}\\
  \langle \eta_{\alpha}(t) \eta_{\beta} (t') \rangle &=& D_{S} \delta_{\alpha \beta} \delta (t - t')
\label{fd_spin_2}
\end{eqnarray}

\subsection{Longitudinal Spin Fluctuation Treatment}\label{app:eom}

\begin{figure}[!tb]
\centering
\includegraphics[keepaspectratio=true, width=3.25in]{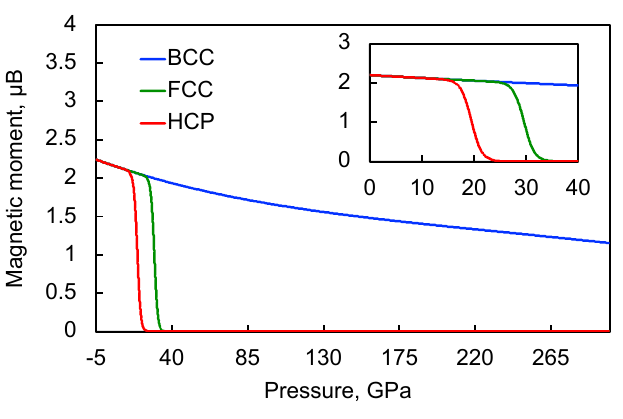}
\caption{Illustration of how magnetic moments vary for BCC, FCC, and HCP phases with pressure. Sigmoidal functions were used to transition the magnetic moments to zero near the BCC-FCC and BCC-HCP transition pressures. The Inset image shows a close-up near BCC-FCC and BCC-HCP transition pressures (x/y axes share the same units as the main plot).}
\label{fig:theoracle}
\end{figure}

Traditional spin-dynamical Fe-modelling efforts in the past, which have examined near ambient regions of phase space, have usually assumed a constant magnetic moment of $\sim2.2\mu B$ \cite{ma2012longitudinal}. This assumption is reasonable at low pressures, within the BCC phase of iron, where the magnetic moment has a weaker dependence on the lattice strain. For the FCC and HCP phases, once the BCC-HCP/FCC transition has been completed the local magnetic moment rapidly drops off to zero~\cite{RBS13}. Hence, in the current shock study of iron, where we compress Fe to Earth-core conditions (300-350 GPa) the standard fixed magnetic moment assumption becomes increasingly problematic. To account for changes in the magnetic moment we implement a longitudinal-spin fluctuation ``fix" (LAMMPS terminology for quantities affecting time integration) within the SPIN package which connects changes of the per-atom magnetic moment with changes in strain and phase within a given atom's local neighborhood. Incorporation of these longitudinal spin fluctuations necessitates the addition of the Landau Hamiltonian, described in Eq.~\ref{HLandau} which describes how changes in spin magnitude alter the magnetic energy.

The use of even powers of the spin vector $S_{i}$ in Eq.~\ref{HLandau}, ensures appropriate adherence to time-reversal symmetry~\cite{ma2012longitudinal}. The vector $S_{i}$, associated with atom $i$, can be defined in terms of the magnetic moment $\bf{M}$ as $S_{i} = -\frac{\bf{M}}{g\mu_{B}}$, where $\mu_{B}$ is the Bohr magneton constant and $g$ is the Lande factor. The coefficients $A$, $B$, and $C$ are defined in Eqs.~\ref{eq:A_coeff}-\ref{eq:C_coeff} shown below. These definitions follow the approach of Ma {\it{et al.}}~\cite{MD20}.

\begin{equation}
A = a(g\mu_{B})^2 \
\label{eq:A_coeff}
\end{equation}
\begin{equation}
B = b(g\mu_{B})^4 \
\label{eq:B_coeff}
\end{equation}
\begin{equation}
C = c(g\mu_{B})^6 \
\label{eq:C_coeff}
\end{equation}

Contrary to Ma {\it{et al.}} however, the definition in Eq.~\ref{eq:A_coeff} assumes that $\mathcal{H}_{ex}$ is offset by the perfect-crystal ground state. To obtain the coefficients $a$, $b$, and $c$ we fit a sixth-order polynomial to $E_{total}$ vs. $M$ data obtained from non-collinear VASP data~\cite{GAEH20}. In doing so, we carry out sixth-order polynomial fits for each $E_{total}$ vs. $M$ curve at discrete values of strain. This allows us to define each of the $a$, $b$, and $c$ coefficients as a function of the local atomic strain (pressure), as shown in Eqs.~\ref{eq:a_coeff}-\ref{eq:c_coeff} below. 
	    
\begin{equation}
\begin{split}
a &= \frac{-0.0273S_{i}(p_{loc})^3 + 0.0789S_{i}(p_{loc})^2}{S_{i}(p_{loc})^2 - 3.457S_{i}(p_{loc}) + 3.284} +  \\
&~ \frac{-0.0757S_{i}(p_{loc}) + 0.018}{S_{i}(p_{loc})^2 - 3.457S_{i}(p_{loc}) + 3.284}
\label{eq:a_coeff}
\end{split}
\end{equation}

\begin{equation}
\begin{split}
b &= 0.006374e^{-\Big[\frac{S_{i}(p_{loc})-2.576}{0.6864}\Big]^2} + \\
&~ 0.008469e^{-\Big[\frac{S_{i}(p_{loc})-0.7717}{2.323}\Big]^2} \
\label{eq:b_coeff}
\end{split}
\end{equation}
\begin{equation}
c = -0.000059 \
\label{eq:c_coeff}
\end{equation}

The sixth-order coefficient $c$ varied very weakly with strain/pressure and was thus, for simplicity, set to a constant (average of all fits). Together with Eqs.~\ref{eq:s_bcc}-\ref{eq:s_hcp}, which define how the local magnetic moment varies within the BCC/FCC/HCP phases of iron with pressure, the relations in Eqs.~\ref{eq:a_coeff}-\ref{eq:c_coeff} effectively allow us to affix changes in the magnetic moment to changes in the local structure (strain and phase). This methodology is adopted within a new LAMMPS "fix" created for this effort, where code is available upon request.
 
\begin{eqnarray}
S_{i,BCC} &= 0.942(4.964\cdot10^{-11}p_{loc}^4 - \nonumber \\
& 6.666\cdot10^{-8}p_{loc}^3 + \nonumber \\ 
& 3.048\cdot10^{-5}p_{loc}^2 - \nonumber \\ 
& 8.192\cdot10^{-3}p_{loc} + 2.335)
                  \label{eq:s_bcc}
\end{eqnarray}

 \begin{eqnarray}
S_{i,FCC} &= S_{i,BCC}(p_{loc}) - \nonumber \\ 
& S_{i,BCC}(p_{loc})\frac{1}{1+e^{-p_{loc} + p_{FCC}}}
\label{eq:s_fcc}
\end{eqnarray}

 \begin{eqnarray}
S_{i,HCP} &= S_{i,BCC}(p_{loc}) - \nonumber \\ 
& S_{i,BCC}(p_{loc})\frac{1}{1+e^{-p_{loc} + p_{HCP}}}
\label{eq:s_hcp}
\end{eqnarray}

In all LSF calculations, we examined a spherical local neighborhood around each atom within a 5 $\AA$ cutoff where the local pressure ($p_{loc}$) and volume are computed. The volumetric calculations were carried out using Voronoi tessellations. Local pressure/strain calculations were used to determine changes in the local magnetic moment as outlined in Eqs.~\ref{eq:a_coeff}-\ref{eq:s_hcp} and were carried out every timestep. Here $p_{FCC}$ and $p_{HCP}$ are the FCC and HCP cold-curve transition pressures, which for the current potential are approx. 25.6 and 15.5 GPa respectively. The corresponding magnetic-moment / pressure relationships for the BCC/FCC/HCP phases of iron are illustrated in Fig.~\ref{fig:theoracle}. For liquid atoms, we make the assumption that the magnetic moment magnitude is identically zero regardless of pressure/volume. 

\begin{figure*}[htp]
\centering
\includegraphics[width=6in]{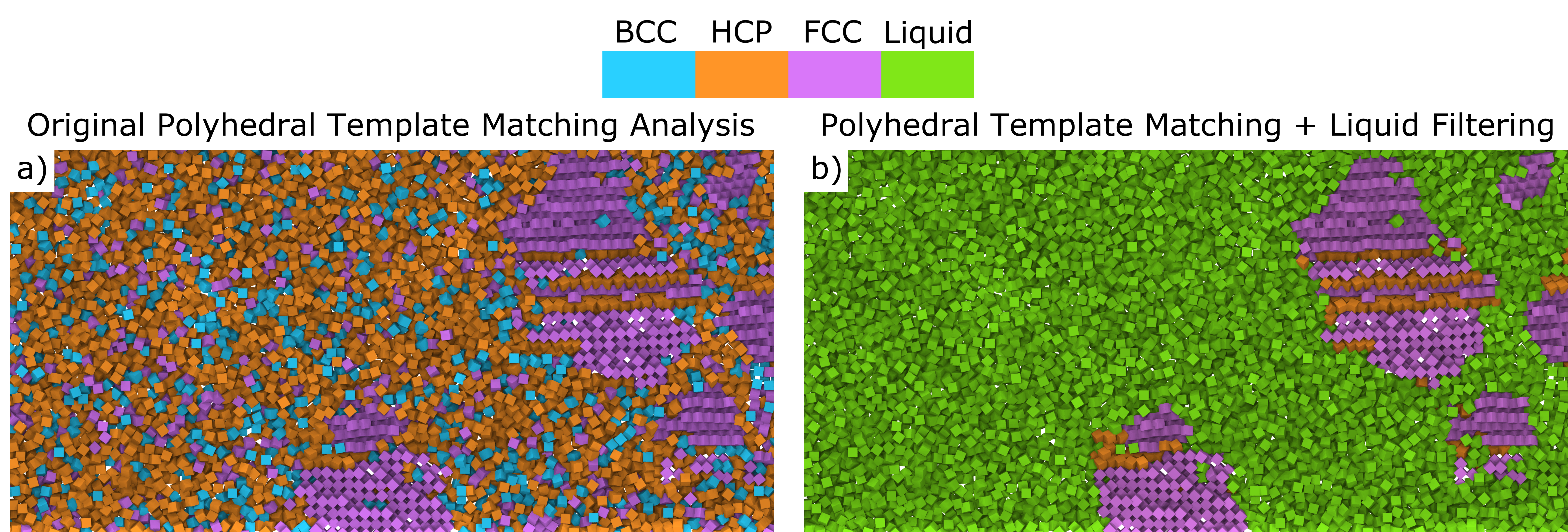}
\caption{a) Section of shocked iron that has undergone partial melting colored by default PTM phase predictions. b) PTM phase predictions after applying the liquid filter. Atoms are displayed as cubes with orientations matching PTM predicted orientation, note random orientation in liquid, and ordered in solid identified phases.}
\label{fig:liquid_detection}
\end{figure*}

Atoms in the liquid phase are identified by using a modified version of the polyhedral-template matching (PTM) fix in LAMMPS because the default PTM method does not correctly identify these states. However, the PTM quaternion vectors provide information regarding the crystal orientation of individual atoms. Thus leveraging the quaternion data and the fact that for liquid regions the PTM predicted crystalline orientations vary significantly, we can unambiguously detect liquid regions within our domain. To achieve this, we modify the standard PTM fix to include a liquid filter, that examines a 5 $\AA$ local neighborhood around each atom. For each pair of atoms within this cutoff, we sum the absolute value of the quaternion dot products, which is then normalized by the number of atom pairs. Using this dot product summation (net orientation similarity) we specify a unique liquid criterion for each atom. For an atom to be reclassified as liquid we consider the default type assigned by PTM (i.e. BCC/FCC/HCP), then assign a local orientation similarity threshold below which said atom is designated as liquid. For atoms identified as BCC/HCP/FCC phases, these thresholds are 0.01, 0.2, and 0.11 respectively. A number density criteria (0.04 atoms/$\AA^3$) is also provided for solid atoms, to account for cases where the default PTM compute may identify a single atom within a liquid region as solid. As shown in Fig.~\ref{fig:liquid_detection} this setup enables a reliable tracking of the solid-liquid domains. In Fig.~\ref{fig:liquid_detection} individual atoms are shown as cubes, where each cube is aligned along the crystalline axes. Thus in solid domains, cubes show similar orientations, whereas in the liquid regions, the cube/atom orientations are random. Note that this liquid filtering does not affect the identification of true solid phases.

\section{Shock simulation details}
For all shock simulations, a momentum-mirror piston is used, where the velocity vector of each atom is reflected at this cell boundary. This fictitious piston wall was moved along the $z$-direction, imposing a fully supported shock wave at velocities $0.1\leq u_{p}\leq 3.5$ km/s. Each simulation geometry was roughly 14.4 x 14.4 x 454.4 nm in size. These are average dimensions as different preheat temperatures cause expansions in bar dimensions. Initial sample temperatures examined within this study were 50$K$, 300$K$, 600$K$, and 1000$K$. Temperature, pressure, and density profiles throughout each bar were taken in 10 $nm$ slices along the $z$-direction and averaged within each sample. Shock speeds ($u_{s}$) are measured by tracking the pressure spike along the $z$-direction. To minimize errors from transient shock response, $u_{s}$ measurements are carried out after the first 20 $ps$ post impact. Each of these settings and geometries is duplicated for both LSFs and fixed-moment (FM) cases.

\section{LSF and FM shock comparison}\label{app:shock_fm_lsf}    

\begin{figure*}[htp]
\centering
\includegraphics[width=\textwidth]{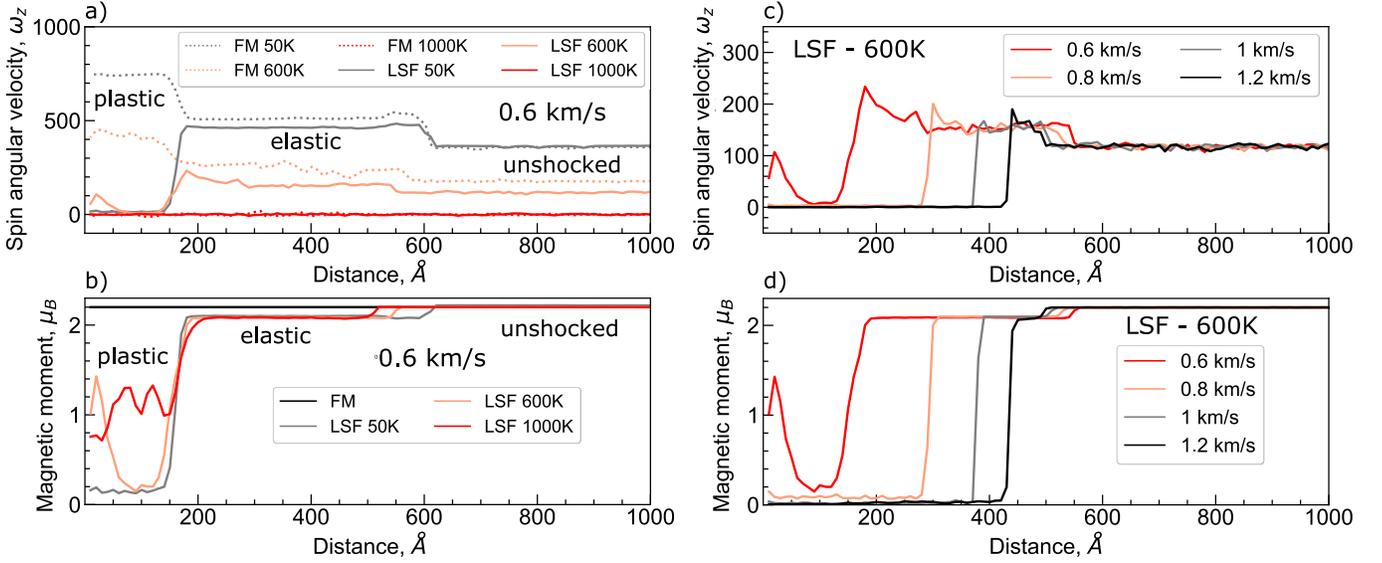}
\caption{ a) Plot showing FM and LSF profiles of the spin angular velocity in the $z$-direction for a 0.6 km/s shock (in $z$-direction), under different preheat temperatures b) Illustration of FM and LSF magnetic moment profiles in the $z$-direction for a 0.6 km/s shock, at different preheat temperatures. c-d) Plots illustrating how the spin angular velocity and magnetic moments vary with different piston velocities for a 600K preheat under the LSF scheme. } 
\label{fig:magplots}
\end{figure*} 

To illustrate the importance of longitudinal spin fluctuations under shock compression we carry out a set of piston impacts up to 1.2 km/s at different preheat temperatures (50-1000K). Each pre-heat will follow a unique path through the (pressure, temperature) phase diagram, this locus of points constrained by total energy conservation as the shock wave passes is called the Hugoniot. These results are summarized in Fig.~\ref{fig:magplots}. Each plot in Fig.~\ref{fig:magplots}.a illustrates how the spin angular velocity in the $z$-direction varies throughout an iron bar subjected to a $u_{p}=0.6$ km/s shock (in $z$-direction), under different preheat temperatures for both treatments of spin magnitudes. For a 1000K preheat temperature, the magnetic disorder is large and no discernible changes in the $\omega_{z}$ signal throughout the bar can be observed. This is true in both FM and LSF cases. At lower temperatures, however, in the FM case, high $\omega_{z}$ values are recorded in the plastically deformed regions of the bar, which retain their ferromagnetic character. This is not observed in the LSF case where ferromagnetism is lost as the magnetic moments decrease due to the HCP phase transformation and corresponding increases in pressure. 

In the elastic front, which is solely made up of a compressed BCC lattice, a small difference between FM and LSF cases is observed. In Fig.~\ref{fig:magplots}.b it is shown how the magnetic moments vary, at different preheats, throughout a bar that has been shocked at 0.6 km/s. In the FM cases, the magnetic moment is always fixed to 2.2 $\mu B$. For the LSF case, in the plastic front, the BCC-HCP transition pressure increases for higher preheat temperatures, increasing the presence of BCC atoms and therefore also the average magnetic moment. In the compressed BCC elastic front, a small decrease in the magnetic moment is observed, from approx. 2.2 to 2 $\mu B$. The plots shown in Fig.~\ref{fig:magplots}.c-d illustrate how the LSF spin angular velocity and magnetic moment profiles change at higher shock speeds (up to 1.2 km/s).
Notably, lower shock speeds (0.6 km/s) do not impart enough energy to fully transform the cell to HCP, and thus turn off all magnetic moments in the plastic wave. In these cases, portions of the plastic front still retain a ferromagnetic character (non-zero $\omega_{z}$). Meanwhile, larger piston velocities increase the length of the plastic region in the shock front.

\section{Elastic property calculations}
For the MSD-LSF elastic property measurements, finite size effects were checked with cubic samples of $16\cdot10^4$ and $2\cdot10^6$ atom counts. For each different domain size, we started with 300K polycrystalline samples at pressures of 140, 180, 240, and 300 GPa. Each was then heated to 5882K and held in the isobaric ensemble at 140, 180, 240, and 300 GPa until melting could be confirmed using pair correlation data. Once melted, equilibration in the micro-canonical ensemble was carried out for 100 ps before compressional wave velocity Eq. \ref{Pwave} was measured.
\begin{equation}
\begin{aligned}
P_{wave} & = \sqrt{\frac{K_S(T) + (4/3)G(T)}{\rho}}
\label{Pwave}
\end{aligned}
\end{equation}
Here $K_S(T)$ is the measured adiabatic bulk modulus and $G(T)$ is the shear modulus which is given by Eq.~\ref{eq:eqG} below. 
\begin{equation}
\begin{aligned}
G(T) & = \frac{7C_{11} -5C_{12} + 2C_{33} - 4C_{13} + 12C_{44}}{30}
\label{eq:eqG}
\end{aligned}
\end{equation}

The finite-temperature elastic constants $C_{11}$, $C_{12}$, $C_{33}$, $C_{13}$, and $C_{44}$ are measured using the born-matrix calculation in LAMMPS, as shown in Eq.~\ref{Born} below.

\begin{equation}
\begin{aligned}
C_{i,j} & = \langle \frac{1}{V}\frac{\partial^2 U}{\partial \epsilon_i \partial \epsilon_j} \rangle + \frac{V}{k_B T}(\langle \sigma_i \sigma_j \rangle -\langle \sigma_i \rangle \langle \sigma_j \rangle ) + \\
& \frac{Nk_BT}{V}(\delta_{i,j} + \sum_{i=1,2,3}\delta_{1,i}\sum_{j=1,2,3}\delta_{2,j} )
\label{Born}
\end{aligned}
\end{equation}

Here $N$, $k_B$, $T$, $\sigma$, $V$, $U$, $\epsilon$, and $\delta$ correspond to the number of particles, Boltzmann constant, temperature, virial stress tensor, volume, potential energy, strain tensor, and Kronecker delta.

\section{Transport property calculations}\label{app:eom}
\begin{figure*}[htp]
\includegraphics[width=7.0in]{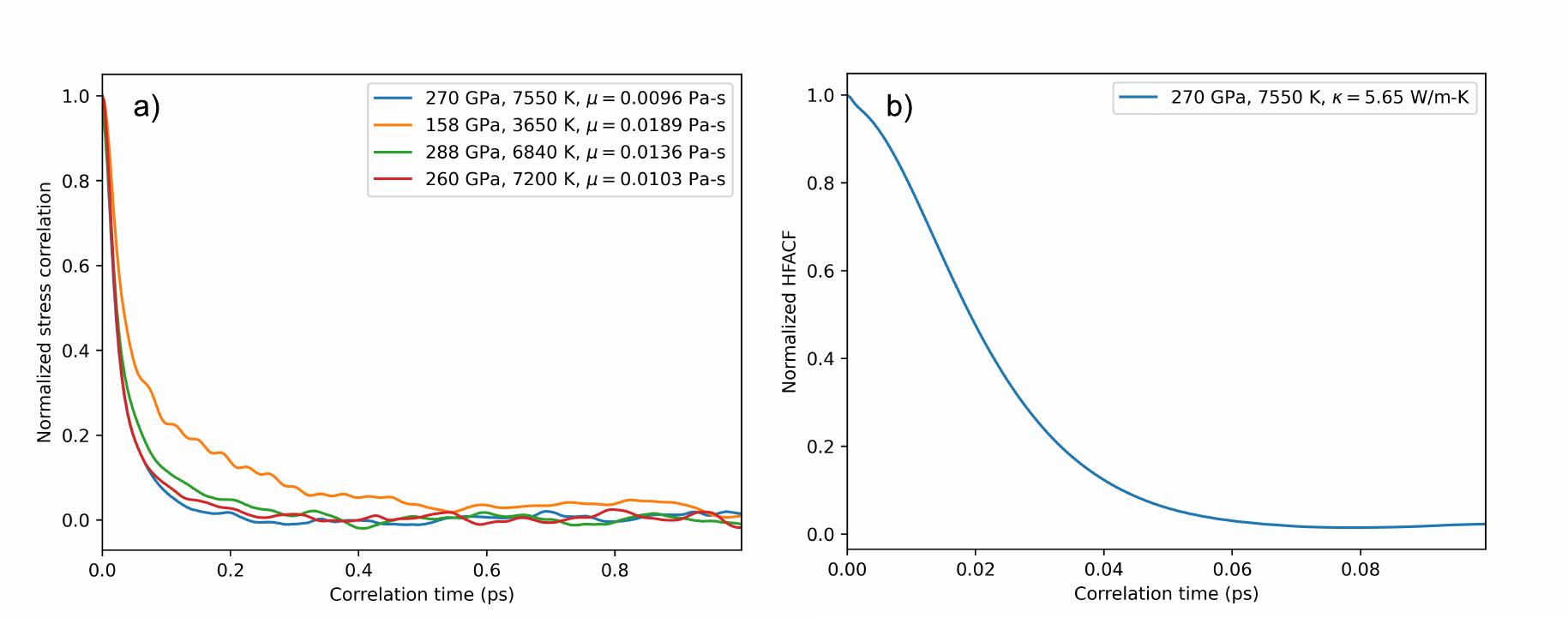}
\caption{a) Normalized shear stress autocorrelation function for different iron liquid states, with their associated viscosity values. These are time-averaged values that had deviations of less than 10 percent over different ensembles. b) Normalized heat flux autocorrelation function for a high-temperature liquid iron state, with its associated thermal conductivity value due to atom motion; note that this value does not include electron contributions.}
\label{fig:autocorrelation}
\end{figure*}

\begin{figure*}[htp]
\centering
\includegraphics[width=7.0in]{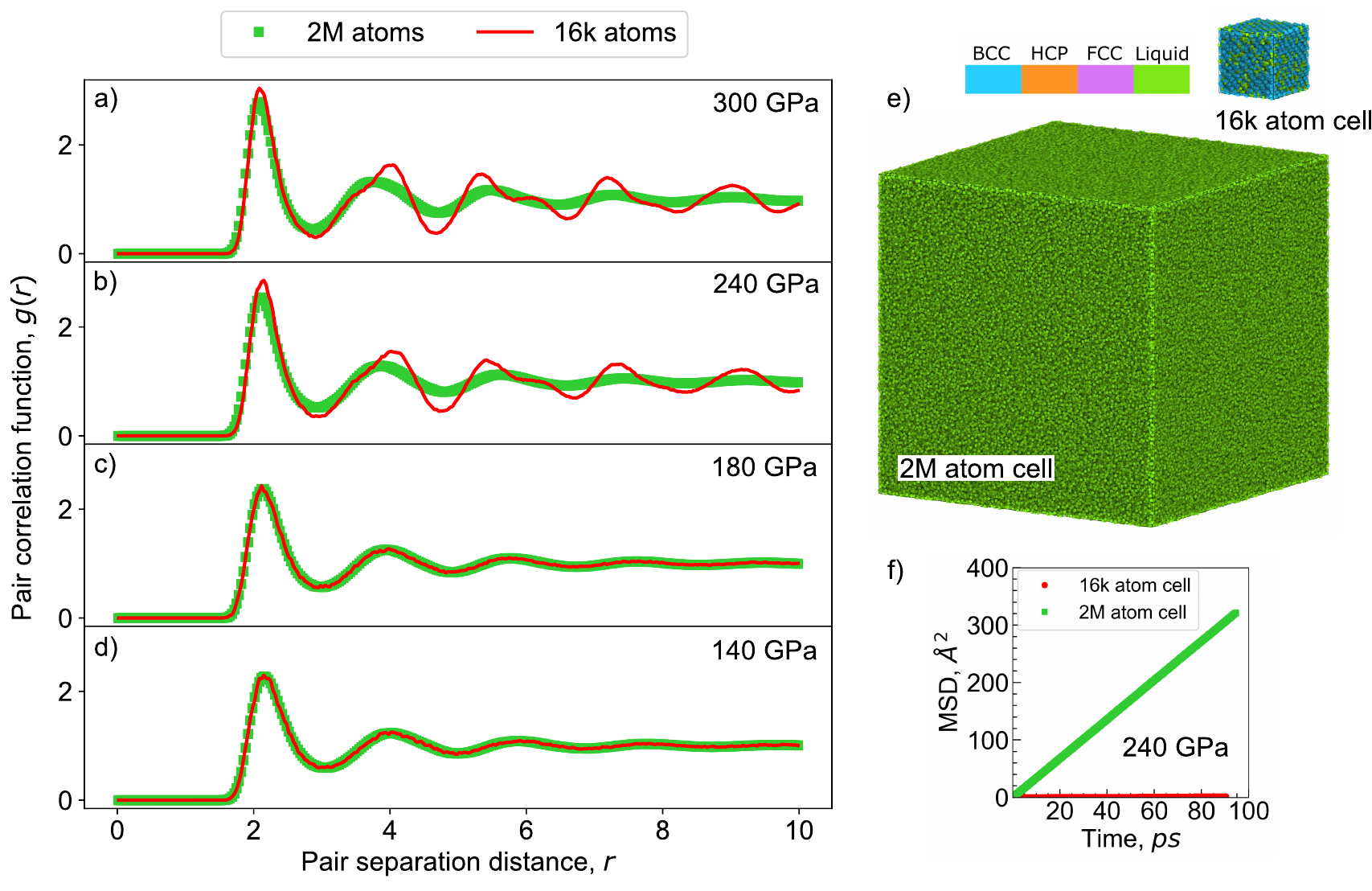}
\caption{a-d) Pair correlation function comparisons between 16k and 2M atom simulation cell cases for 5882$K$ isotherm. Plots a-b highlight the persistence of metastable BCC phase in smaller cell sizes at high pressures. e) Snapshots at the end of 100 $ps$ NVE equilibration (post thermostating) illustrating 16k and 2M atom domains. As can be seen by the legend, the 16k atom cell seems to stabilize the BCC phase better compared to the 2M equivalent. f) Mean-square displacement (MSD) comparison at 240 GPa for 16k and 2M cases. Fluid behavior for the 2M atom case is observed, whereas the 16k atom cell behaves as a solid.} 
\label{fig:metastable_bcc}
\end{figure*} 

\begin{figure*}[htp]
\centering
\includegraphics[width=5.0in]{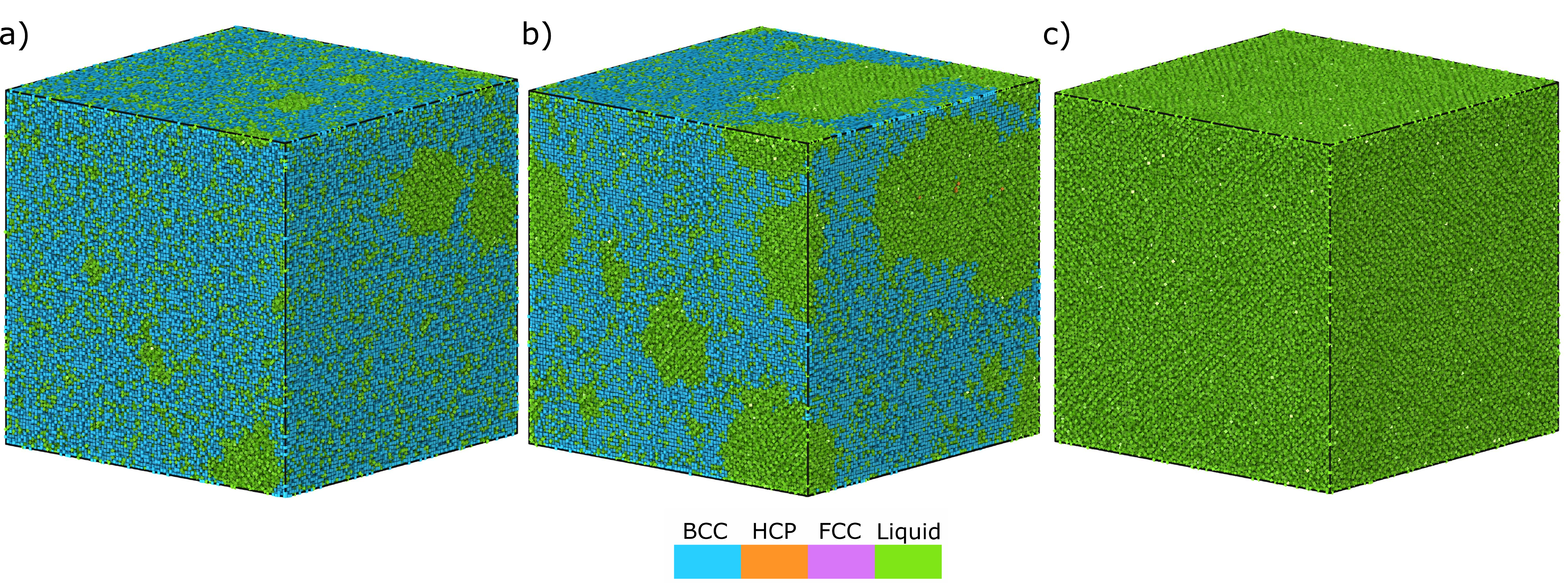}
\caption{a-c) Snapshots over 100 $ps$ time-span illustrating how melting is initiated from high pressure BCC phase.} 
\label{fig:bcc_melt}
\end{figure*} 

By spatially sub-sampling the high temperature/pressure shock data, which contains millions of atoms, we can explore how important transport properties, like thermal conductivity and viscosity, vary at Earth-core conditions. Spatial sectioning of our shock simulation data allows us to extract smaller chunks of material, on the order of 10,000 atoms, from the liquid domains. Note these smaller volumes are representative of the final shocked state, but cannot be considered as free from finite size/time restrictions if used to capture the loading path and transient behavior. This will be explained further by highlighting metastable phases near the melt line. Using the Green-Kubo method~\cite{G54,K57} (based on linear response theory) we then measure the phonon thermal conductivity and viscosity from classical MSD. We find that the phonons contribute minimally to the thermal conductivity at these high temperatures/pressures (~1 Wm/K). For the Earth-core liquid states, the viscosity shows a higher sensitivity to the motion of atoms, since thermal conductivity is largely dominated by electronic contributions under these conditions.

The linear response of shear stress is studied via the Green-Kubo (GK) formula, which for viscosity $\mu$ is given by time integrals of autocorrelation functions of the shear stress signal $P_{ab}(t)$, where the indices $a$ and $b$ can be any Cartesian component $x,y,z$. This formula for viscosity is given by:
      
\begin{equation}
\mu_{ab} = \frac{V}{k_B T} \int_0^{\infty} \langle P_{ab}(0) P_{ab}(t) \rangle dt \
\label{eq:GK_viscosity}
\end{equation}
        
This formula is evaluated in practice by performing time and/or ensemble averages over MD trajectory data, where the integral is evaluated up to a point where the autocorrelation function decays, i.e. decorrelated atomic motion. We found that a variety of high-temperature liquid states exhibited similar autocorrelation decay times, as shown in Fig.~\ref{fig:autocorrelation}.a.
        
As seen in Fig.~\ref{fig:autocorrelation}.a the shear stress autocorrelation decays to near zero within 1 ps; we, therefore, time-integrated Eq.~\ref{eq:GK_viscosity} out to 1 ps. This integration was performed in many time windows throughout 1 ns simulations and checked for consistency by performing 3 different ensembles of such calculations, whereby the predicted values deviated by less than 10 percent. Similar quickly decaying autocorrelation functions were seen with the atom heat flux, as shown in Fig.~\ref{fig:autocorrelation}.b. Similarly to viscosity, we calculate atomic motion contributions to thermal conductivity from the Green-Kubo formula:
        
\begin{equation}
\kappa_{ab} = \frac{V}{k_B T} \int_0^{\infty} \langle Q_{ab}(0) Q_{ab}(t) \rangle dt \
\label{eq:GK_thermal}
\end{equation} 
where $Q_{ab}(t)$ is the atomic heat flux signal. 
As seen in Fig. \ref{fig:autocorrelation}.b, the heat flux becomes uncorrelated with itself after 0.06-0.08 ps. We therefore perform the integration in Eq.~\ref{eq:GK_thermal} out to 0.1 ps. After performing this time integration the resulting phonon thermal conductivity of the 270 GPa 7550 K liquid iron state is determined to be 5.65 W/mK. This phonon thermal conductivity does not include electronic contributions, however, which are dominant still in liquid metals.

\begin{figure*}[htp]
\centering   
\includegraphics[width=6.5in]{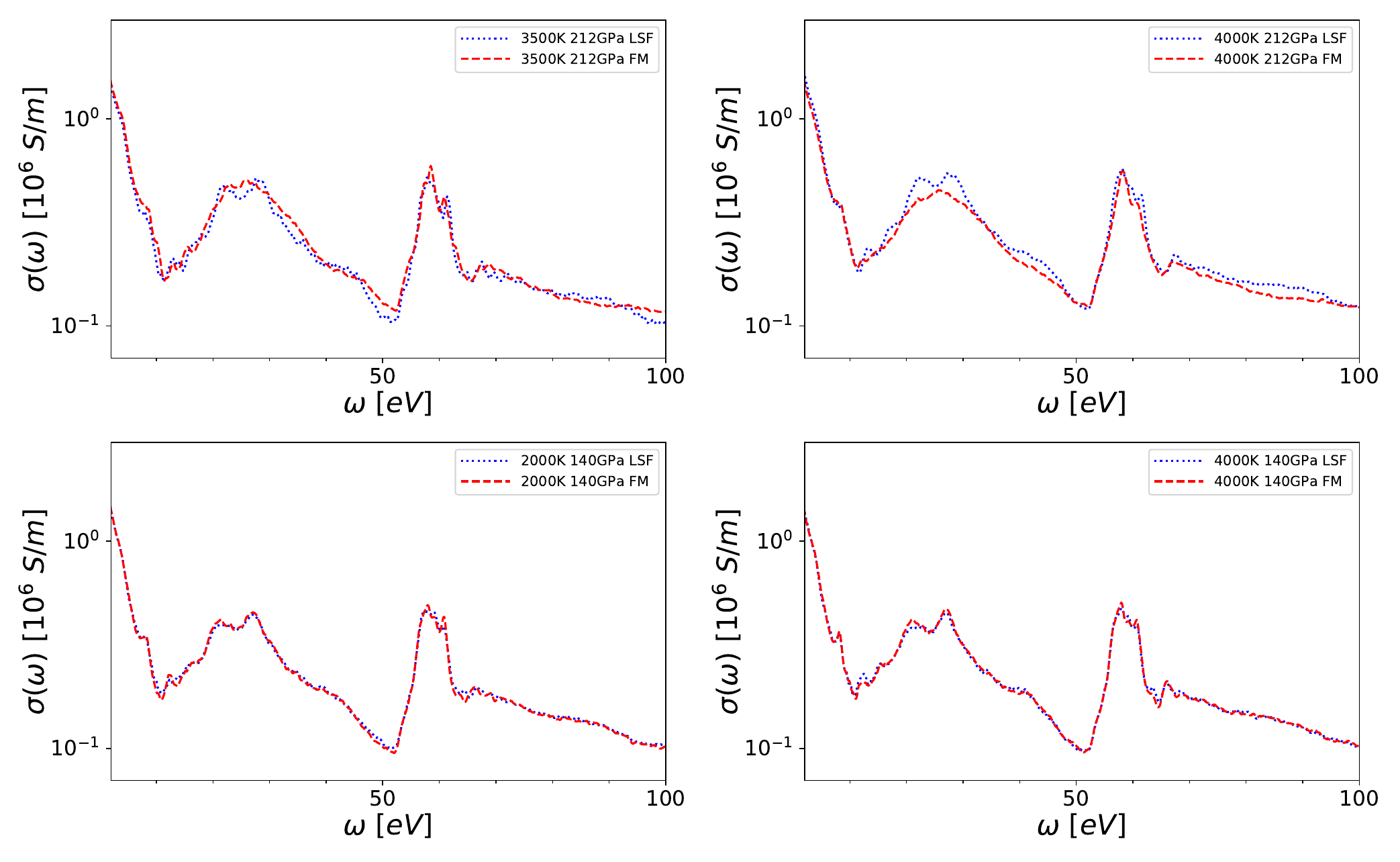}
\caption{\raggedright Frequency-dependent electrical conductivity evaluated using RT-TDDFT on configurations obtained from shock simulations for LSF and FM.} 
\label{supp:sigma_comp_fm_lsf} 
\end{figure*}

To highlight the impact of finite-size effects on the melt-boundary we examined the pair correlation functions for both 16k and 2M atom cells for a 5882 $K$ isotherm within the pressure range of 140-300 $GPa$. We note that in all cases we observe the presence of a metastable BCC phase that is a precursor to melting. However, as shown in Fig.~\ref{fig:metastable_bcc}, above 200 GPa the smaller 16k atom cells remain stuck within these metastable configurations for much longer periods compared to the larger 2M atom cells. Importantly we note that in Fig.\ref{fig:metastable_bcc}.f the MSD for the 16k atom cell is linearly increasing in time just at a significantly slower rate than the 2M case. In Fig.~\ref{fig:bcc_melt}, provided snapshots illustrate how melting proceeds from the BCC phase in a 2M atom cell. These results underpin the challenges of properly resolving melt-transition boundaries at high pressures, which are particularly prohibitive for ab-initio methods, where access to large spatial/temporal scales is not possible.

Electrical and thermal conductivity at high pressures and temperatures relevant to the conditions of the Earth's interior is dominated by the electronic component. Here, the electrical conductivity is computed using the real-time formalism of time-dependent density functional theory (RT-TDDFT). RT-TDDFT~\cite{YB96} and the methodology to extract the electrical conductivity are described in Refs.~\cite{RLBV23,Ramakrishna2023iop}. The RT-TDDFT results are obtained from an all-electron full-potential linearized augmented plane wave (FP-LAPW) method~\cite{SN06} as implemented in the Elk~\cite{D22} and Exciting~\cite{GKMN14,PD21} codes. Note, that we use Hartree atomic units everywhere in the parameters for TDDFT calculations. Within the RT-TDDFT calculations, a sigmoidal pulse of vector amplitude 0.1~a.u. is applied with a peak time of 2~a.u. with a full-width half maximum (FWHM) of 0.5~a.u. for a total simulation time up to 1000~a.u (1 a.u.=24.819 attoseconds) providing energy resolution up to 7~mHa in frequency space. The calculations were performed on a \emph{k}-point grid consisting of $4\times 4\times 4$ points using PBE~\cite{PBE96}  exchange-correlation functional. The DC conductivity is extracted from a Drude-like fit to a restricted set of low-energy spectra. The atomic configurations for the input of RT-TDDFT simulations are generated using LSF/FM MSD simulations in LAMMPS. Figure~\ref{supp:sigma_comp_fm_lsf} shows the dynamical electrical conductivity based on RT-TDDFT calculations for configurations obtained using LSF/FM MSD simulations at shocked conditions of P=212~GPa and P=140~GPa, and, temperature ranging from 2000~K to 4000~K evaluated for a system size consisting of N=16 atoms. 
                
\section{\textit{Ab-initio} Molecular Dynamics Training Data}
\label{app:kushal-vasp}

\begin{figure}[htp]
\centering
\includegraphics[width=1.0\columnwidth]{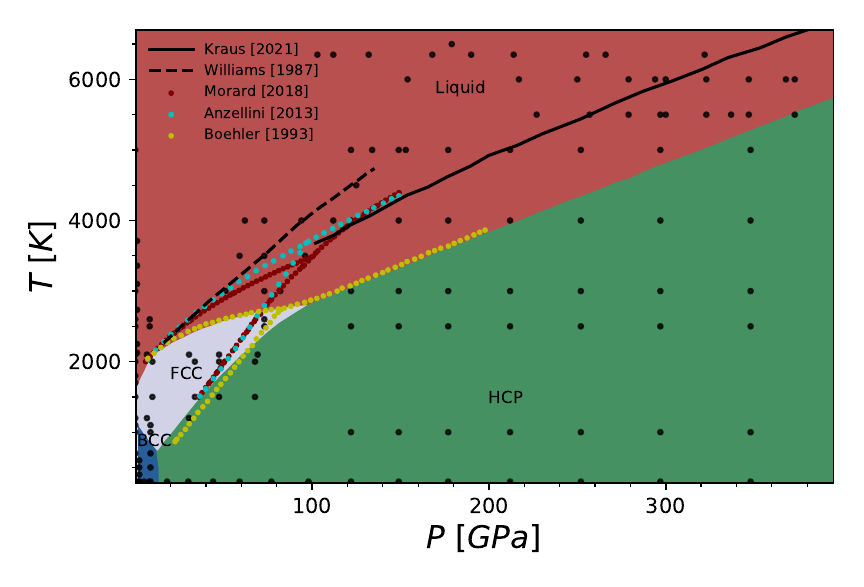}
\caption{\raggedright Phase diagram of iron from Ref.~\onlinecite{THOT10} indicated in shaded regions. Phase boundaries and the melting curve obtained from experiments are additionally shown~\cite{B93,ADML13,MBRA18,KHAB22,WJBS87}. The \textit{ab-initio} simulations considered for training in this work are indicated by black markers juxtaposed to the phase diagram from Ref.~\onlinecite{THOT10}.  } 
\label{fig:phase}
\end{figure}  

Training data for ML-MSD model is obtained from density functional molecular dynamics (DFT-MD) simulations, which are performed using VASP~\cite{KH93,KF96,KF96b,KJ99}. PAW pseudopotentials~\cite{B94} with a Perdew-Burke-Ernzerhof (PBE)~\cite{PBE96} exchange-correlation (XC) functional with 16 valence electrons and a core radius of $r_c=1.9 a_B$ are used. 
The plane wave cutoff was set to 750~eV and the convergence in each self-consistency cycle was set to $10^{-5}$. 
We verified energy cutoff up to ~1000~eV with a cutoff of 600~eV and above ensuring the total energy deviation is $<$0.1$\%$.  
We used the Mermin formulation of thermal DFT~\cite{M65} and Fermi occupation of the eigenvalues. Generally, the first Brillouin zone was sampled on a $2\times 2\times 2$ grid of \textit{k}-points in an N=16 supercell, and at the gamma-point in an N=128 supercell. The number of bands varied up to 300 and 1600 for the highest temperature ($\sim$6500~K) and pressures up to $\sim$430~GPa in an N=16 and N=128 supercell respectively. The thermostat in the NVE ensemble was of Nose-Hoover type~\cite{H85}. Ionic time steps of $\Delta t = 0.2$~fs were taken. 
Simulations were carried out on various phases (BCC/FCC/HCP/Liquid) in the pressure-temperature plane of the iron phase diagram~\cite{THOT10,B93,ADML13,MBRA18,KHAB22,WJBS87} shown in Fig.~\ref{fig:phase}.

\bibliography{si.bib} 